\documentclass[superscriptaddress,aps,pre,twocolumn,a4paper]{revtex4}

\usepackage{graphicx}
\usepackage{ulem}
\usepackage{amssymb}
\usepackage[utf8x]{inputenc}
\usepackage{amsmath}
\usepackage{listings}
\DeclareMathOperator{\sech}{sech}

\begin{document}

\title {Improved transfer matrix methods for calculating quantum transmission-coefficient}
 
\vspace{1.2in}
\author{Debabrata Biswas}
\affiliation{Theoretical Physics Division, Bhabha Atomic Research Centre, Mumbai 400085}
\author{Vishal Kumar}
\affiliation{Centre for Excellence in Basic Science, Mumbai University, Kalina, Mumbai 400 098}

\begin{abstract}
Methods for calculating the transmission coefficient are proposed, all of which arise
from improved non-reflecting  WKB boundary conditions at the edge of the 
computational domain in 1-dimensional geometries. In the first,
the Schr\"{o}dinger equation is solved numerically while the second
is a transfer matrix (TM) algorithm where the potential is approximated by steps, but
with the first and last matrix modified to reflect
the new boundary condition. Both methods give excellent results with first order
WKB boundary conditions. The third  uses the transfer matrix method 
with third order WKB boundary conditions. For the the parabolic potential, the 
average error for the modified third order TM method reduces by factor 
of 4100 over the unmodified TM method.
\end{abstract}

\date{\today}

\maketitle
\newcommand{\be}{\begin{equation}}
\newcommand{\ee}{\end{equation}}
\newcommand{\bea}{\begin{eqnarray}}
\newcommand{\eea}{\end{eqnarray}}
\newcommand{\Tbar}{{\bar{T}}}
\newcommand{\ep}{{\cal E}}
\newcommand{\Lop}{{\cal L}}
\newcommand{\DB}[1]{\marginpar{\footnotesize DB: #1}}
\newcommand{\q}{\vec{q}}
\newcommand{\kt}{\tilde{k}}
\newcommand{\Lopn}{\tilde{\Lop}}
\newcommand{\noi}{\noindent}
\newcommand{\ovn}{\bar{n}}
\newcommand{\ovx}{\bar{x}}
\newcommand{\ovE}{\bar{E}}
\newcommand{\ovV}{\bar{V}}
\newcommand{\ovU}{\bar{U}}
\newcommand{\ovJ}{\bar{J}}
\newcommand{\calE}{{\cal E}}
\newcommand{\ovphi}{\bar{\phi}}
\newcommand{\oveps}{\bar{\ep}}

\section{Introduction}

The transmission coefficient in quantum mechanics relates the probability flux carried by the
transmitted wave relative to the incident wave. It is used in tunneling calculations
such as field-emission from metals \cite{FN,DB14}, quantum cascade lasers \cite{JK2014}
or more generally when dealing with electron transport at 
the nanoscale \cite{vasileska,nature11,cassan,Mao}. 
An accurate and computationally effective method to determine this quantity is thus
desirable.

In a wide variety of situations where the tunneling region is thin, 
a 1-dimensional modeling of the tunneling process is adequate. As an example, 
ultra-thin oxide barriers in metal-oxide-semiconductor (MOS) devices
can be modeled using a single degree of freedom. The methods developed 
in this paper to improve the
accuracy of transmission coefficient calculation are
of relevance in such quasi 1-dimensional systems.

The WKB formula \cite{landau} for transmission coefficient (TC)

\be
TC = \frac{e^{-{2\over \hbar} \int_{x_1}^{x_2} \sqrt{2m(E - V(x))} dx}}
{(1 + \frac{1}{4} e^{-{2\over \hbar} \int_{x_1}^{x_2} \sqrt{2m(E - V(x))} dx})^2}
 \label{eq:WKB}
\ee

\noi
in 1-dimensional systems
is the most widely used one in literature. Here $x_1,x_2$ are the two
classical turning points at an energy $E$ and $V(x)$ is the 
potential energy. Eq.~(\ref{eq:WKB}) is applicable for energies 
less than the barrier height when tunneling
occurs in position space. An analogous WKB formula for above-barrier tunneling
can be derived in momentum space at least for simple potentials \cite{pWKB,landau}. 
While these
formulae are easy to use and reasonably accurate at energies for which barriers 
are broad and high, they are inappropriate for tunneling near the top of the barrier
or above-barrier reflection from generic potentials where the momentum space tunneling
formula may be hard to implement.

It is thus necessary to rely on numerical methods to determine the 
transmission coefficient, either by solving the time-independent Schr\"{o}dinger 
equation explicitly (referred to hereafter as Differential Equation or DE method)
or by approximating the
potential by a series of steps or line-segments and using the transfer matrix (TM) formalism.
The step-approximation TM method \cite{TM,CJ2009} is one of the most widely used 
numerical schemes. It is simple to use since the matrix elements are known
analytically and it only requires $N$ matrices to be multiplied where $N$ is the degree of 
discretization. Normally, convergence is obtained rapidly with a few thousand matrices.

Both the DE and TM methods   mentioned 
above have an approximation 
in common \cite{holds_elsewhere,continued_fraction}. Since numerical methods require
a finite domain, they require boundary conditions. 
This essentially implies that a form
for the potential must be assumed beyond the computational domain that is easy to
solve so that the wavefunction and its derivative may be matched at the boundary.
Normally, it is assumed that the potential is constant beyond the computational
domain so that plane wave solutions exist. This allows both the DE and TM methods to be specified
fully. For the Schr\"{o}dinger equation approach (DE), appropriate boundary conditions can be
derived while for the transfer matrix method, the boundary matrices can be
determined. The results in both cases are generally better than the WKB formula.
Our aim here is to go beyond the plane wave assumption mentioned above to provide 
a non-reflecting truncation scheme for the computational domain and test
it by calculating the transmission coefficient.

In Section \ref{sec:boundary1}, we first review the standard approximation involved in
truncating boundaries and then go beyond plane waves by using first order WKB wavefunctions.
This is used to derive new boundary conditions for solving the time-independent
Schrodinger equation as well as new transfer matrices at the boundary.
The first order boundary conditions are implemented numerically in 
section \ref{subsec:numerics} using potentials for
which the exact transmission coefficients are known. In Section \ref{sec:boundary3},
we provide the formalism for third order WKB boundary conditions and 
implement the same using transfer matrices. Our results are summarized in
section \ref{sec:conclusions}.

\section{Boundary Truncation using first order WKB}
\label{sec:boundary1}

As mentioned above, a finite computational domain requires boundary conditions that
allow flux to be transmitted without causing spurious reflections. In 1-dimensional
situations, it is generally accepted that this can be achieved by
assuming that the flux beyond is carried away by plane waves. This 
essentially implies that the potential assumes a constant value on either side of the 
computational domain. The discontinuity in the first derivative of the potential
however requires a reflected wave from the boundary in order that the wavefunctions
and their first derivatives match. To see this, let the computational domain be $L \leq x \leq 0$
with a boundary at $x = 0$. For values of $x$ slightly
less than zero, the potential may be approximated by $V(x) = V(0) - \alpha x$
where  $\alpha = -V'(0)$.
Thus, the Schr\"{o}dinger equation takes the form

\be
-\frac{\hbar^2}{2m} \frac{d^2}{dx^2}\psi(x) - \alpha x \psi(x) = \ep \psi(x)
\ee

\noi
where $\ep=E-V(0)$. 
For $\ep > 0$, the solutions are 

\be
\psi(x) = \left\{
\begin{array}{c} 
\sqrt{z} H_{1/3}^{(1)}(2z^{3/2}/3) \\  
\sqrt{z}H_{1/3}^{(2)}(2z^{3/2}/3) \end{array} \right.
\ee

\noi
where $z = (\alpha x + \ep)(2m)^{1/3}(\alpha \hbar)^{-2/3}$ and $H_{1/3}^{(1,2)}$ are
Hankel functions.
A general solution in the computational domain near the boundary at $x = 0$ can
thus be expressed as 

\be
\psi_-(x) = C \sqrt{z} H_{1/3}^{(1)}(2z^{3/2}/3) + D \sqrt{z}  H_{1/3}^{(2)}(2z^{3/2}/3)
\ee

\noi
where $ H_{1/3}^{(1)}$ represents
a wave moving to the right and $H_{1/3}^{(2)}$ a reflected wave moving to the left
from the computational domain. Matching $\psi_-(x)$ and its derivative
to the plane wave solution for $x \geq 0$, $\psi_+(x) = F \exp(ikx)$,  
leads to a solution where
both $C$ and $D$ are non-zero. Thus, the plane wave assumption leads to 
spurious reflection, its magnitude depending on the factors such as the
energy $\ep = E - V(0)$.

There is thus scope to improve upon this truncation scheme. One possibility is
to assume that the wavefunction at the end of the computational domain is
a first order semiclassical WKB wavefunction

\be
\psi_+^{wkb}(x) = \frac{F}{\sqrt{p(x)}} e^{\frac{1}{\hbar}\int^x p(x') dx'}  \label{eq:wkb1}
\ee

\noi
which can be matched at the boundary. The lower limit in the integral
in Eq.~(\ref{eq:wkb1}) is an appropriately chosen reference point.  
We shall build upon this approach
first proposed in the context of the self-consistent Schrodinger-Poisson 
system \cite{epjb,epl}. Here, as in the plane 
wave case, it is assumed that there is no reflection from beyond the
computational domain so that a left moving wave is not included in $\psi_+^{wkb}(x)$.
In addition, it is also assumed that the end of the computational domain
is not a classical turning point for the energy considered and that the
potential is sufficiently slowly varying over a deBroglie wavelength.

\subsection{Improved boundary conditions for the Schrodinger Equation (DE method)}

For purposes of determining the transmission coefficent, it is easier to
write the wavefunction in polar form

\be
\psi(\ovx) = (\frac{J m D}{e\hbar})^{1/2}~~r(\ovx)e^{i\theta(\ovx)} \label{eq:psi_exact}
\ee

\noi
where $\ovx = x/D$, $D$ is the extent of the computational domain and
$r(\ovx)$ and $\theta(\ovx)$ are real valued functions. 
For convenience in writing the Schr\"{o}dinger equation in dimensionless form,
it is assumed that the tunneling particle is
an electron with charge $-e$ and mass $m$. Thus, the tunneling current
density is 

\be
J  = \frac{ ie\hbar}{2m} ( \psi^* \frac{d\psi}{dx} - \psi \frac{d\psi^{*}}{dx}) \label{eq:current}.
\ee

\noi
The Schr\"{o}dinger equation thus reduces to equations for the amplitude and phase: 

\bea
&~& \frac{d^2 r}{d\ovx^2} + [(\ovE - \ovV) - \frac{1}{r^4} ]r = 0 \label{eq:Sch}\\
&~& \frac{d \theta}{d\ovx} = \frac{1}{r^2}
\eea

\noi
where $\ovV = V/(eV_s)$, $\ovE = E/(eV_s)$ and $V_s = \hbar^2/(2meD^2)$. 
For simplicity, we shall assume the left computational boundary to be at $x = 0$
and the right boundary at $x = D$ or $\ovx = 1$.

Note that
once $r(\ovx)$ is known, $\theta(\ovx)$ can be determined independently 
with an arbitrary phase ($\theta(1) = 0$) at the right boundary $\ovx=1$.
The boundary conditions for $r(\ovx)$ are easier to implement at $\ovx=1$.
It is thus simpler to 
solve Eqn. (\ref{eq:Sch}) as an initial value problem
starting at $\ovx=1$.
Our task thus reduces to the determination of $r(1)$ and $r'(1)$.

Using Eqns.~(\ref{eq:wkb1}) and  (\ref{eq:current}), the real coefficient $F$ can
be expressed in terms of the current density $J$ as $F = \sqrt{mJ/e}$.
On matching the wavefunctions and their derivatives at $\ovx = 1$, we have

\bea
r(1) & = & \frac{1}{(\ovE - \ovV(1))^{1/4}} \label{eq:ini1}\\
r'(1) &  = & - \frac{\ovV'(1)}{[4(\ovE - \ovV(1) )^{5/4}]} \label{eq:ini2}.
\eea

\noi
In contrast, for a plane outgoing wave, $r'(1) = 0$.
Eq.~(\ref{eq:Sch}) can be integrated backward to determine $r(0)$ and $r'(0)$.
The transmission
coefficent can be obtained by 
matching $\psi(\ovx)$  to the WKB form 

\be
\psi_-^{wkb}(x) =  \frac{A}{\sqrt{p(x)}} e^{\frac{1}{\hbar}\int^x p(x') dx'} + 
\frac{B}{\sqrt{p(x)}} e^{-\frac{1}{\hbar}\int^x p(x') dx'}
\ee

\noi
at $x = 0$. The transmission coefficient can thus be expressed in terms of
$r(0)$, $r'(0)$ and $\ovV'(0)$ as 
 
\be
{\cal T} = \frac{4}{(\frac{r(0)\ovV'(0)}{4\oveps^{5/4}} + \frac{r'(0)}{\oveps^{1/4}})^2
+ (\oveps^{1/4} r(0) + \frac{1}{\oveps^{1/4} r(0)})^2}
\ee

\noi
where $\oveps = (\ovE - \ovV(0))$. An analogous expression for plane outgoing
waves can be obtained for comparison.

\subsection{First Order WKB Transfer Matrices}

Solving the Schr\"{o}dinger equation explicitly using WKB boundary conditions
improves computation of the transmission coeffient as we shall see in the next
section. Here, we shall investigate whether the popular transfer matrix method
can be tweaked to incorporate the WKB truncation technique.

In the transfer matrix method, instead of dealing with a continuous variation of potential and
solving the differential equation,
the potential is divided into several segments (steps).
Each of these segments behaves as an individual potential step and since the
segments are small, the potential is considered to be constant for a given segment.
The potential tends towards the actual value as the divisions becomes finer.
The open boundary is treated by WKB aproximation assuming the 
potential to be slowly varying outside the computational domain.

\begin{figure}[htb]
 \centering
 \includegraphics[width=6cm,angle=270]{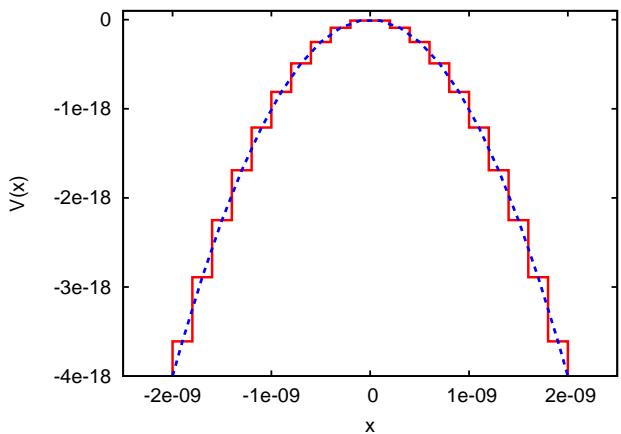}
 \caption{The potential energy $V(x) = -x^2$ (dashed line) approximated by steps.}
\label{fig:step}
\end{figure}

Let us consider a potential $V(x)$ which is divided into $N$ segments as shown
in Fig.(1) and let the computational domain be $[x_0,x_{N}]$. 
Thus, in Fig.~\ref{fig:step}, $x_0 = -2$nm and $x_{N} = 2$nm.
The potential is approximated by a multistep function

\begin{equation}
 V(x)= V_j=V[(x_{j-1} +x_j)/2]
\end{equation}

\noi
for $ x_{j-1} < x < x_j~$, $j=1,2,\ldots,N$.
The wave function $\psi_{j}$ in the $j_{th}$ region for an electron
with energy $E$ is given by $ \psi_j(x)=A_j e^{ik_jx} + B_je^{-ik_jx}$
where $k_j= \sqrt{2m(E-V_j)}~/\hbar$ for $j=1,\ldots,N$ and $\hbar = h/2\pi$, $h$ being Planck's constant.
Consider a  first order WKB wavefunction
to the left of the computational domain 

\begin{equation}
\psi_0^{wkb}(x \leq x_0)=\frac{A_0}{\sqrt {\hbar k_0}} e^{i \int^{x} k(x')dx'} + \frac{B_0}{\sqrt {\hbar k_0}} e^{-i \int^{x} k(x')dx'}
\end{equation} 

\noi
with $k_0 = \sqrt{2m(E-V(x_0))}~/\hbar$
and a plane wave as the wavefunction for the first step of the potential 

\begin{equation}
\psi_1(x)=A_1 e^{i k_1x} + B_1 e^{-i k_1x}.
\end{equation}

\noi
Applying continuity of the wavefunction and its derivative at the boundary $x = x_0$ we get,

\begin{equation}
\begin{split}
\left( \frac{A_0}{\sqrt {\hbar k_0}} e^{i \int^{x} k(x')dx'} + \frac{B_0}{\sqrt {\hbar k_0}} e^{-i \int^{x} k(x')dx'} \right)_{{x=x_0}} \\
= \left( A_1 e^{i k_1x} + B_1 e^{-i k_1x} \right)_{{x=x_0}}
\end{split}
\end{equation}

\noi
and 

\begin{equation}
\begin{split}
\left[ \frac{A_0}{\sqrt {\hbar k_0}}  \left( ik_0 - \frac{1}{2} \frac{k'_0}{k_0}  \right) e^{i \int^{x} k(x')dx'} \right]_{{x=x_0}}  + \\
\left[ \frac{B_0}{\sqrt {\hbar k_0}} \left( -ik_0 -\frac{1}{2} \frac{k'_0}{k_0} \right) e^{-i \int^{x} k(x')dx'} \right]_{{x=x_0}} \\
= ik_1 \left( A_1 e^{i k_1x} - B_1 e^{-i k_1x} \right)_{{x=x_0}} .~~~~~~~~~~~~~~
\end{split}
\end{equation}

\noi 
Choosing the reference point (lower limit) for the integration to be $x_0$ itself, $\int^{x_0} k(x')dx' = 0$.
Thus, from the above equations one can write the transfer matrix for the left boundary as

\be
M_0 = {\frac{1}{2\sqrt{\hbar k_0}}} \left[ \begin{array}{cc} (1+\gamma_{0}^+) e^{-i k_1 x_0} &
(1-\gamma_{0}^-) e^{-i k_1 x_0} \\ (1-\gamma_{0}^+) e^{i k_1x_0} &
(1+\gamma_{0}^-) e^{ik_1x_0}  \end{array} \right]
\ee

\noi
where $\gamma_{0}^- = \alpha_{0}-\beta_{0}$, $\gamma_{0}^+ = \alpha_{0} + \beta_{0}$, 
$\alpha_{0} = k_0/k_1$ and $\beta_{0} = ik'_0/(k_0k_1)$ with 
$k'_0=-m V'(x_0)/(k_0 \hbar^2)$.

From the continuity equations at the boundaries of succesive segments, the value of $A_j$
and $B_j$ can be reduced to a multiplication of the $j$ ($2\times2$) matrices

\be
\left( \begin{array}{c} A_j \\ B_j \end{array} \right) 
=\prod_{l=0}^{j-1} M_l \left( \begin{array}{c} A_0 \\ B_0 \end{array} \right)
\ee

\noindent
where

\be
M_l=  \frac{1}{2} \left[ \begin{array}{cc} (1+\alpha_l) e^{-i(k_{l+1} - k_l)x_l} &
 (1-\alpha_l) e^{-i(k_{l+1} + k_l)x_l} \\
(1-\alpha_l) e^{i(k_{l+1} + k_l)x_l} & (1+\alpha_l) e^{i(k_{l+1} - k_l)x_l}  \end{array} \right]
\label{eq:genericM}
\ee

\noi
where
$\alpha_l= k_l/k_{l+1}$, $l = 1, \ldots, N-1$. At the right end of the computational domain (i.e. $x = x_{N}$), 
the WKB wavefunction takes the form

\begin{equation}
\begin{split}
\psi_{N+1}(x=x_{N})=\frac{A_{N+1}}{\sqrt {\hbar k_{N+1}}} e^{i \int^x k(x')dx'} \\
 + \frac{B_{N+1}}{\sqrt {\hbar k_{N+1}}} e^{-i \int^x k(x')dx'}  
\end{split}
\end{equation}

\noi
with $ k_{N+1} = \sqrt{2m(E-V(x_N))}~/\hbar$.
For $x_{N-1} \leq x \leq x_{N}$, the wavefunction is given by

\begin{equation}
\psi_N(x)=A_N e^{i k_Nx} + B_N e^{-i k_Nx}.
\end{equation}

\noi
On applying continuity equations  at the right boundary $x = x_{N}$, it follows that

\begin{equation}
\begin{split}
\left( A_N e^{i k_Nx} + B_N e^{-i k_Nx} \right)_{{x=x_N}} = ~~~~~~~~~~~~~~~~ \\
\left( \frac{A_{N+1}}{\sqrt {\hbar k_{N+1}}} e^{i \int^x k(x')dx'}
 + \frac{B_{N+1}}{\sqrt {\hbar k_{N+1}}} e^{-i \int^x k(x')dx'} \right)_{{x=x_N}}
\end{split}
\end{equation} 

\noi
and 

\be
\begin{split}
ik_N \left( A_N e^{i k_Nx} - B_N e^{-i k_Nx} \right)_{{x=x_N}} = ~~~~~~~~~~~~~\\
\left[ \frac{A_{N+1}}{\sqrt {\hbar k_{N+1}}} e^{i \int^{x} k(x')dx'} \left( ik_{N+1} - 
\frac{1}{2} \frac{k'_{N+1}}{k_{N+1}} \right)\right]_{x=x_N} + ~~ \\ 
\left[ \frac{B_{N+1}}{\sqrt {\hbar k_{N+1}}} e^{-i \int^{x} k(x')dx'} \left( -ik_{N+1} -\frac{1}{2} \frac{k'_{N+1}}{k_{N+1}} 
\right)\right]_{{x=x_N}}.
\end{split}
\end{equation} 

\noi
The phase factors $e^{\pm i \int^{x} k(x')dx'}$ can be absorbed in the coefficents $A_{N+1}$ and $B_{N+1}$
since their absolute value determines the transmission coefficient. Thus,
from the above continuity equations, one can write the transfer matrix, $M_{N}$, for the right boundary as 

\be
\frac{\sqrt{\hbar}}{2i\sqrt{k_{N+1}}}\left[ \begin{array}{cc} (iS_N^+ + R_N) e^{i k_N x_N} &
(iS_N^- + R_N) e^{-i k_N x_N} \\
(iS_N^- - R_N) e^{i k_N x_N} &
(iS_N^+ - R_N ) e^{-i k_N x_N}  \end{array}\right]
\ee

\noi
where  $S_N^+ = k_{N+1} + k_N$, $S_N^- = k_{N+1} - k_N$
$R_N = k_{N+1}'/2k_{N+1}$ and  $k'_{N+1} = -m V'(x_N)/(k_{N+1} \hbar^2)$.

Let us consider the amplitude of the incident wave $A_0 = 1$  and the final reflected wave $B_{N+1}= 0$.
Therefore the transmission amplitude $A_{N+1}$ is given as

\bea
A_{N+1} & = & \frac{\det(M_0)\det(M_1)...\det(M_N)}{M_{22}} \\ 
& = & \frac{k_1}{k_N}\frac{\det(M_0)\det(M_N)}{M_{22}} \label{eq:An}
\eea

\noi
where
\be
M=\left( \begin{array}{c} M_{11}~~~~~M_{12} \\ M_{21}~~~~~M_{22} \end{array} \right) 
=\prod_{k=0}^{N} M_k.
\ee 

\noi
The transmission probability $TC$ is thus

\begin{equation}
TC = \mid A_{N+1} \mid^2.
\end{equation}

\subsection{Numerical Results}
\label{subsec:numerics}

\begin{figure}[tbh]
 \centering
 \includegraphics[width=6cm,angle=270]{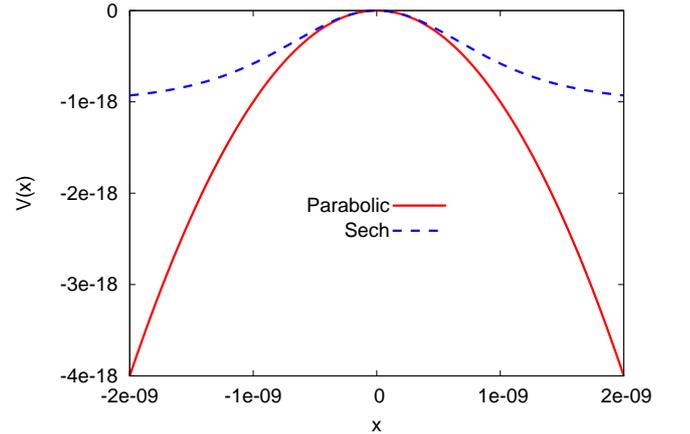}
 \caption{The Parabolic potential, $V(x) = -x^2$ (solid line) and the secant 
hyperbolic (Sech)  potential 
$V(x) = V_0 (\sech^2(x/x_0) - 1)$ (dashed line) within the computational domain [-2nm,2nm].
Here $V_0 = 1\times10^{-18}$J and $x_0 = 1\times10^{-9}$m. }
\label{fig:pots}
\end{figure}

\begin{figure}[bth]
 \centering
 \includegraphics[width=6cm,angle=270]{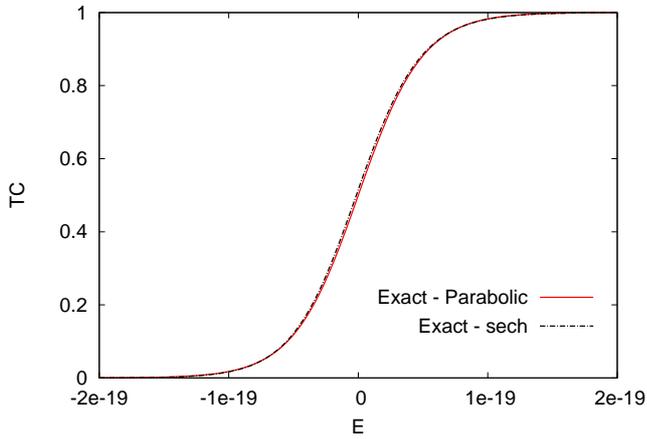}
 \caption{Transmission coefficient (TC) for the parabolic (solid) and secant hyperbolic 
potentials (dashed).
The energy range covers the variation of TC from 0 to 1. }
\label{fig:exact}
\end{figure}

\begin{figure}[tbh]
 \centering
 \includegraphics[width=6cm,angle=270]{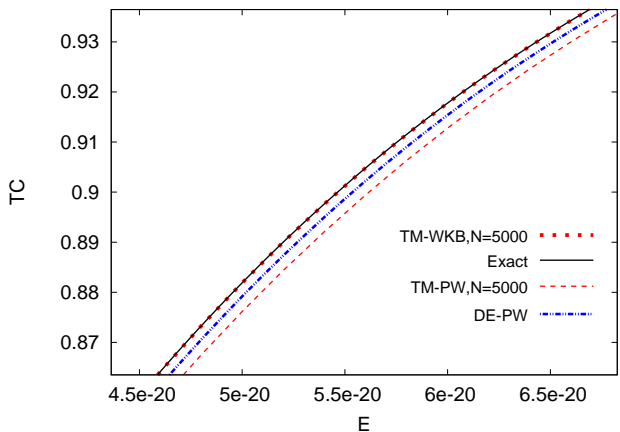}
 \caption{A comparison of transmission coefficient obtained by four numerical methods
along with the exact result for the parabolic potential $V(x) = -x^2$. The DE 
method using WKB boundary condition is hard to distinguish from the 
exact or TM-WKB results and is therefore not shown. The transfer matrix methods are labelled TM. 
The DE-PW (middle curve) fares better than TM-PW. The
solid line is the exact result.}
\label{fig:parabolic}
\end{figure}

We present results for two potentials $V(x) = -x^2$ and $V(x) = V_0 (\sech^2(x/x_0) - 1)$,
$V_0 = 1\times10^{-18}$J, $x_0 = 1\times10^{-9}$m (see Fig.~\ref{fig:pots}), for
which the exact transmission coefficients are known \cite{landau} 
(see Fig. \ref{fig:exact}). The computational
domain used is from [-2nm,2nm] while the energy range over which the transmission coefficient
varies from 0 to 1 is $-2\times10^{-19}$J to $2\times10^{-19}$J. The two potentials are
shown in Fig.~(\ref{fig:pots}). While the parabolic potential rapidly decreases 
for $|x| > 0$, the slope of the secant hyperbolic potential decreases for increasing $|x|$ 
with a saturation value $V(\pm\infty) = -1 \times 10^{-18}$. It can thus be expected
that with a computational domain $|x| \leq 2 \times 10^{-9}$m, the plane wave method for the 
secant hyperbolic potential
should fare reasonably well alongside the WKB methods. Unless otherwise mentioned, all 
distances plotted are in metres and energy (including potential $V(x)$) in joule.

In Fig.~(\ref{fig:parabolic}), we present a comparison of the numerical methods discussed along with 
the exact result. Two of these use the differential equation (DE) approach where the
Schr\"{o}dinger equation is solved but with 
plane and WKB waves respectively at the boundary of the computational domain. The other two
are the transfer matrix methods (TM), again with plane and WKB waves at the boundary
of the computational domain. 
The TM and DE methods with WKB boundary condition are clearly the best (DE-WKB is not shown in the figure 
since it is indistinguishable from TM-WKB and the exact result). The Plane Wave (PW) methods have
errors with DE-PW better than TM-PW method.

\begin{figure}[tbh]
 \centering
 \includegraphics[width=6cm,angle=270]{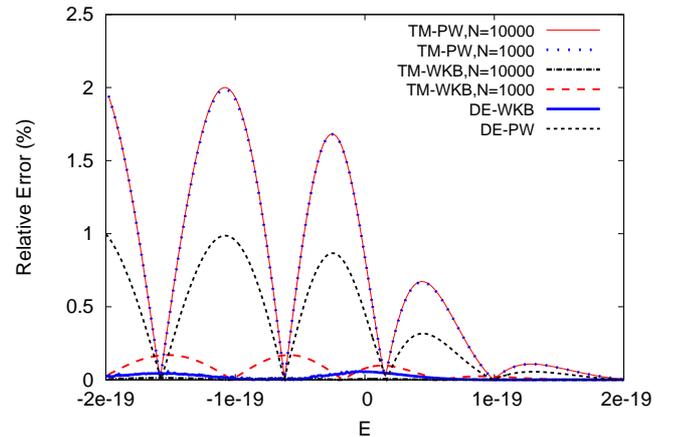}
 \caption{Relative error in transmission coefficient for the parabolic potential for the
four numerical methods discussed above. The largest errors are in the TM-PW method (top two)
followed by the DE-PW method. }
\label{fig:error1}
\end{figure}

\begin{figure}[tbh]
 \centering
 \includegraphics[width=6cm,angle=270]{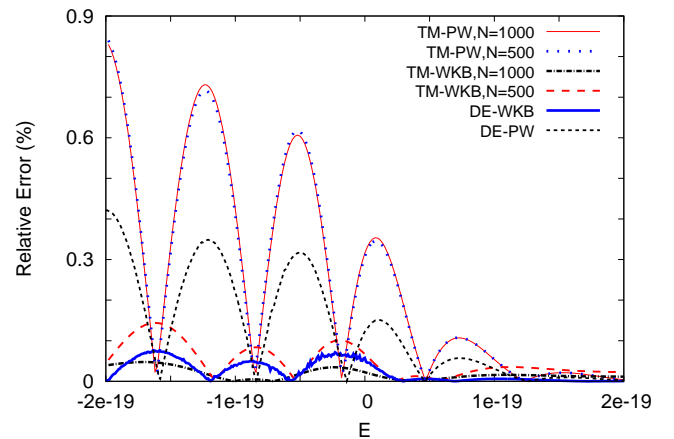}
 \caption{Relative error in transmission coefficient for the secant hyperbolic potential
as in Fig.~\ref{fig:error1}. }
\label{fig:error2}
\end{figure}

The improvement with WKB boundary conditions in both the DE and TM methods however
depends on the energy under consideration. We have thus computed the relative error
using the exact result for both potentials. These are plotted in Figs.~(\ref{fig:error1})
and (\ref{fig:error2}) for the parabolic and secant hyperbolic potentials respectively.

\begin{figure}[tbh]
 \centering
 \includegraphics[width=6cm,angle=270]{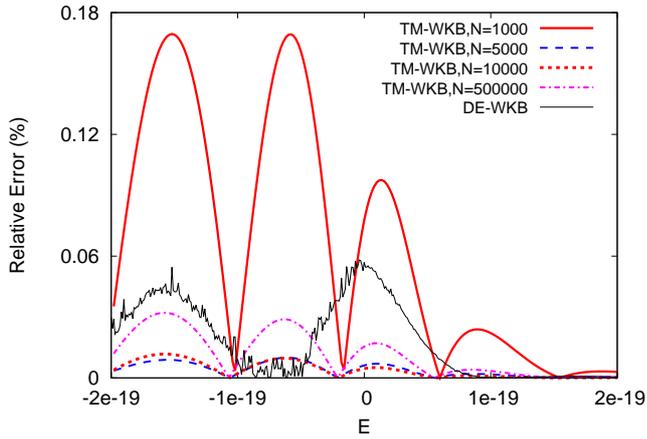}
 \caption{Relative error in transmission coefficient for the parabolic potential for TM-WKB method
with different values of $N$. Also shown is the DE-WKB result (irregular solid curve) for comparison.}
\label{fig:paraN}
\end{figure}

Note that for the TM-PW method, the error saturates fast with $N$ for both potentials
whereas for the TM-WKB method, the error is seen to reduce with $N$. 
Also, both the TM-WKB and DE-WKB methods perform much better than the plane wave counterparts
at all energies with the TM method outperforming the DE method for 
a few thousand steps ($N$). Note that
for the secant hyperbolic potential, a reduced computational domain (for instance $|x| \leq 1$nm), 
leads to a greater
improvement for the WKB methods over the plane-wave methods as expected.

Finally, we compare the error in the TM-WKB method as the number of steps, $N$, is
increased. This is shown in Fig.~\ref{fig:paraN} for the parabolic potential.
The error reduces initially as $N$ is increased but beyond $N=5000$, the error grows slightly 
before saturating at around $N=10^5$ to values that are still lower than the DE-WKB method. 
The energy-averaged error for the saturated TM-WKB method is about 65 times less
than the saturated TM-PW method for $N=10^5$.

\section{Higher Order WKB }
\label{sec:boundary3}

In the previous section, a first order WKB wavefunction was used to determine
non-reflecting boundary conditions at the edge of the computational domain
resulting in considerable improvement of accuracy in the transmission
coefficient.

The method can be easily generalized to achieve higher order WKB boundary
conditions. We shall, however, restrict outselves to third order WKB wavefunctions
in much of what follows and show that the transfer matrix formalism can
be modified further to achieve orders of magnitude improvement in 
accuracy over the first order TM results.

On using the usual WKB expansion  for the wavefunction $\psi(x) = e^{\frac{i}{\hbar} S(x)}$ 
with $S(x) =  \sum_{n=0}^{\infty} \hbar^n S_n $, the Schr\"{o}dinger equation
yields the following equations for $S_n'(x)$:

\bea
S'_0 & = & \pm p(x) \\
S'_1 & = & \frac{i}{2} \frac{p'(x)}{p(x)} 
\eea

\bea
S'_2 & = & \pm \left[ -\frac{p''(x)}{4p^2(x)} + \frac{3(p'(x))^2}{8p^3(x)} \right] \\
S'_3 & = & -\frac{ip'''(x)}{8p^3(x)} + \frac{i3p'(x)p''(x)}{4p^4(x)} - \frac{i3(p'(x))^3}{4p^5(x)}
\label{eq:Sn}
\eea

\noi
where $p(x) = \sqrt{2m(E-V(x))}$. The first two terms $S_0(x) = \pm \int^x p(x') dx'$
and $S_1(x) = (i/2) \ln|p(x)|$ give rise to the first order WKB wavefunction
used in the previous section.

Note the terms are alternately real and imaginary. Thus $S_0$ and $S_2$ give
rise to a phase while $S_1$ and $S_3$ contribute to the amplitude. Further,
$S_0$ abd $S_2$ can assume positive or negative values depending on the 
sign of the momentum $p(x)$. A standard right moving can thus be expressed
as $\psi_+(x) = e^{\frac{i}{\hbar}S_i(x)}$ while a left moving wave can be expressed
as $\psi_-(x) = e^{\frac{i}{\hbar}S_r(x)}$ where

\bea
S_i(x) & = & +S_0(x) + \hbar S_1(x) + \hbar^2 S_2(x) + \hbar^3 S_3(x) \\
S_r(x) & = & -S_0(x) + \hbar S_1(x) - \hbar^2 S_2(x) + \hbar^3 S_3(x).
\eea

A general third order WKB wavefunction at the left end of the 
computational domain is thus

\be
\psi_0(x)=A_0 e^{iS_i(x)/ \hbar} + B_0 e^{iS_r(x)/ \hbar}.
\ee

\noi
In the transfer matrix formalism, the matrix $M_0$ is thus

\be
M_0 = \left[ \begin{array}{cc} 
(1+\frac{S'_i}{\hbar k_1}) e^{i(  \frac{S_i}{ \hbar} - k_1 x_0)} &
(1+\frac{S'_r}{\hbar k_1}) e^{i(  \frac{S_r}{ \hbar} - k_1 x_0)} \\
(1-\frac{S'_i}{\hbar k_1}) e^{i(  \frac{S_i}{ \hbar}+k_1 x_0)} &
(1-\frac{S'_r}{\hbar k_1}) e^{i(  \frac{S_r}{ \hbar}+k_1 x_0)}
 \end{array} \right] \label{eq:M0}
\ee

\noi
where $S_i$, $S_r$, $S_i'$ and $S_r'$ are evaluated are $x_0$.
Similarly, the wavefunction at the right end is

\be
\psi_{N+1}(x)=A_{N+1} e^{iS_i(x)/ \hbar} + B_{N+1} e^{iS_r(x)/ \hbar}
\ee

\noi
so that the transfer matrix $M_{N}$ is

\be
M_N = \frac{1}{S_i' - S_r'} \left[ \begin{array}{cc} 
(\hbar k_N-S_r')e^{i\alpha_i^+} &
-(\hbar k_N+S_r')e^{i\alpha_i^-} \\
-(\hbar k_N-S_i')e^{i\alpha_r^+} &
(\hbar k_N+S_i')e^{i\alpha_r^-}
\end{array} \right]  \label{eq:MN}
\ee

\noi
where $\alpha_i^+ = +k_Nx - S_i/\hbar$, $\alpha_i^- = -k_Nx - S_i/\hbar$,
$\alpha_r^+ = +k_Nx - S_r/\hbar$ and $\alpha_r^- = -k_Nx - S_r/\hbar$.
Here $S_i$,$ S_r$, $S_i'$ and $S_r'$ are evaluated are $x_N$.

As before, with $A_0 = 1$ and $B_{N+1} = 0$, the amplitude

\be
A_{N+1} = \frac{k_1}{k_N} \frac{\det(M_0) \det(M_N)}{M_{22}}
\ee

\noi
with $M = \prod_{l=0}^N M_l$, where $M_0$ and $M_N$ are given by Eqns.~(\ref{eq:M0})
and (\ref{eq:MN}) respectively  while for other values of $l$, $M_l$ is given by
Eq.~(\ref{eq:genericM}). 

\begin{figure}[tbh]
 \centering
 \includegraphics[width=6cm,angle=270]{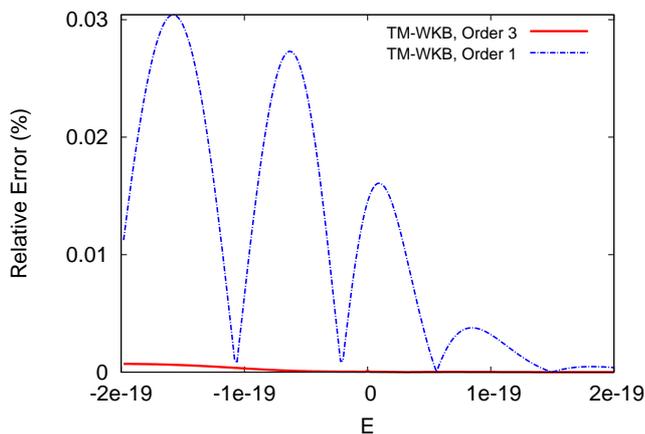}
 \caption{Relative error in transmission coefficient for the parabolic potential using
the transfer matrix method with $1^{st}$ (dashed) and $3^{rd}$ (continuous) order WKB boundaries.}
\label{fig:ho}
\end{figure}

The incident and transmitted currents can be expressed respectively as

\bea
J_{IN} & = & \frac{|A_0|^2}{2m} 2\Re(S_i'(x_0))   e^{-\frac{2}{\hbar} \Im(S_i(x_0))}  \\
J_{TR} & = & \frac{|A_{N+1}|^2}{2m}  2\Re(S_i'(x_N))  e^{-\frac{2}{\hbar} \Im(S_i(x_N))} 
\eea

\noi
where $\Re$ and $\Im$ denote the real and imaginary part respectively. The transmission 
coefficient $TC = J_{IN}/J_{TR}$ is thus

\be
TC  =   |A_{N+1}|^2 \frac{\Re(S_i'(x_N))}{\Re(S_i'(x_0))}  e^{\frac{2}{\hbar}(\Im(S_i(x_0)) - \Im(S_i(x_N)))} 
\ee

\noi 
since $A_0$ = 1.

In order to check whether higher order terms improve the accuracy of the transmission
coefficient, we consider the parabolic potential $V(x) = -x^2$. For convenience,
we consider the reference point for integrating Eqns.~(\ref{eq:Sn}) as the left
boundary ($x = x_0$) of the computational domain so that $S_n(x_0) = 0$ for $n = 0,1,2,3$.
Using $p(x) = \sqrt{2m(E + x^2)}$,  Eqns.~(\ref{eq:Sn}) can be integrated to obtain 
$S_n(x_N)$.

In  Fig.~\ref{fig:ho}, the transmission coefficient obtained using first and third order
WKB boundary conditions are compared for $N = 10^5$ at which both results converge \cite{earlier}. 
Our results are shown in Fig.~\ref{fig:ho}. The energy averaged improvement in relative error
over the first order WKB result is 63 times while the average  improvement over the
plane wave method is about 4100 times. 

\vskip 0.0025 in
$\;$

\section{Summary and Conclusions}
\label{sec:conclusions}

We have demonstrated that the use of WKB wavefunctions at the boundary of 
the computational domain improves the evaluation of the transmission coefficient
enormously. For the parabolic potential, the error reduces by a factor of
4100 using third order transfer matrix method over the usual plane wave
TM method.

It is important to note that the errors are largest at lower energies.
This has significance in field emission calculations where
the supply function may have large contributions below the
Fermi level. The transfer matrix method with WKB boundary condition
(TM-WKB) may thus be adopted due to the ease of implementation
and the improvement in accuracy. Finally, the method can be 
directly generalized to multi-dimensional systems when the 
potential is separable.

\vskip -0.25 in
$\;$

\section{Acknowledgements}

The authors acknowledge stimulating discussions with Dr.~Raghwendra Kumar.



\begin{thebibliography}{99}
\bibitem{holds_elsewhere} The approximation is also used in the continued fraction 
method \cite{continued_fraction} for
determining the transmission coefficient.
\end{thebibliography}
\end{document}